\documentclass[%
reprint,
superscriptaddress,
 amsmath,amssymb,
 aps,
]{revtex4-2}

\usepackage{graphicx}% Include figure files
\usepackage{dcolumn}% Align table columns on decimal point
\usepackage{bm}% bold math
\usepackage{hyperref}% add hypertext capabilities
\usepackage{multirow}
\usepackage[mathlines]{lineno}% Enable numbering of text and display math
\usepackage{color}
\usepackage{ulem}

\begin{document}

\preprint{APS/123-QED}

%\title{Search for sub-GeV invisible particles in the process of $J/\psi\rightarrow\phi + \textrm{invisible}$}
\title{Search for sub-GeV invisible particles in inclusive decays of $J/\psi$ to $\phi$}

\author{
%\begin{small}
\begin{center}
%% Saved at => 2024-12-25
M.~Ablikim$^{1}$, M.~N.~Achasov$^{4,c}$, P.~Adlarson$^{77}$, X.~C.~Ai$^{82}$, R.~Aliberti$^{36}$, A.~Amoroso$^{76A,76C}$, Q.~An$^{73,59,a}$, Y.~Bai$^{58}$, O.~Bakina$^{37}$, Y.~Ban$^{47,h}$
% ...
\\
\vspace{0.2cm}
(BESIII Collaboration)\\
\vspace{0.2cm} {\it
$^{1}$ Institute of High Energy Physics, Beijing 100049, People's Republic of China\\
% ...
$^{4}$ Budker Institute of Nuclear Physics SB RAS (BINP), Novosibirsk 630090, Russia\\
% ...
$^{36}$ Johannes Gutenberg University of Mainz, Johann-Joachim-Becher-Weg 45, D-55099 Mainz, Germany\\
$^{37}$ Joint Institute for Nuclear Research, 141980 Dubna, Moscow region, Russia\\
% ...
$^{47}$ Peking University, Beijing 100871, People's Republic of China\\
% ...
$^{58}$ Southeast University, Nanjing 211100, People's Republic of China\\
$^{59}$ State Key Laboratory of Particle Detection and Electronics, Beijing 100049, Hefei 230026, People's Republic of China\\
% ...
$^{73}$ University of Science and Technology of China, Hefei 230026, People's Republic of China\\
% ...
$^{76}$ University of Turin and INFN, (A)University of Turin, I-10125, Turin, Italy; (B)University of Eastern Piedmont, I-15121, Alessandria, Italy; (C)INFN, I-10125, Turin, Italy\\
$^{77}$ Uppsala University, Box 516, SE-75120 Uppsala, Sweden\\
% ...
$^{82}$ Zhengzhou University, Zhengzhou 450001, People's Republic of China\\
\vspace{0.2cm}
$^{a}$ Deceased\\
% ...
$^{c}$ Also at the Novosibirsk State University, Novosibirsk, 630090, Russia\\
% ...
$^{h}$ Also at State Key Laboratory of Nuclear Physics and Technology, Peking University, Beijing 100871, People's Republic of China\\
% ...
}
\end{center}
%\vspace{-0.2cm}
%\end{small}
}

%\collaboration{BESIII Collaboration}%\noaffiliation

\date{\today}% It is always \today, today,

\begin{abstract}

A search for an invisible particle, $X$, with a mass between 0 and 0.96 $\textrm{GeV}/\textit{c}^{2}$, is performed in the process $J/\psi\rightarrow\phi + X$ using $(8774.0\pm39.4)\times10^{6}$ $J/\psi$ events collected with the BESIII detector from 2017 to 2019.
The $\phi$ meson is fully reconstructed and an efficient veto of photons, neutral and charged hadrons up to twice the $K_L^0$ mass is applied to the rest of the event and the recoil mass against the $\phi$ is obtained precisely from the kinematic constraint in the event.
No significant signal over the expected background is observed in the investigated region and the upper limit on the inclusive branch fraction of $J/\psi\rightarrow\phi + X$ is determined to be $7.0\times10^{-8}$ at 90\% confidence level.
Upper limits at a 90\% confidence level are also given for this branch fraction as a function of the invisible particle mass, varying from $4\times10^{-9}$ to $4\times10^{-8}$ over the investigated mass range. Additionally, a 90\% confidence level upper limit on the branch fraction of $\eta\rightarrow \rm{invisible}$ is determined to $2.4\times10^{-5}$, which improves the previous
best results by more than four times. The analysis technique in this work offers a clean window to search for sub-GeV invisible particles, which can be adapted for other $J/\psi$ decays and direct $e^+e^-$ annihilation experiments in future studies, and improve the sensitivity by orders of magnitude.
\end{abstract}

\maketitle

Despite the strong evidence of dark matter (DM) from astrophysics and cosmology \cite{Corbelli:1999af, Cooper:2009kx, Bertone:2004pz},
DM is still not observed in accelerator experiments after decades of searches \cite{Bertone:2016nfn}. The light DM in sub-GeV (MeV-GeV) range has been widely investigated in both theoretical and experimental efforts \cite{krnjaic2022snowmass} in the last decades, including many studies in intensity-frontier experiments.
Due to the weak coupling with standard model (SM) particles and the light mass scale,
light DM candidates in the sub-GeV range, such as dark photons \cite{Holdom:1985ag, Caputo:2021eaa} and axion-like particles (ALPs) \cite{McKay:1978wn, Jaeckel:2010ni}, are difficult to observe at high energy colliders and by traditional direct detection methods \cite{krnjaic2022snowmass}.
The $e^+e^-$ colliders have unique advantages where sub-GeV light DM candidates behave as invisible particles due to their weak coupling with SM particles, and their kinematics can be inferred from the conservation of four-momenta when the remaining particles in an event are reconstructed.

Previous searches associated to sub-GeV invisible particles have been performed using many data samples from high intensity facilities \cite{BESIII:2023jji, Belle-II:2023ueh, Belle-II:2022jyy, Belle-II:2019qfb, BESIII:2020sdo, BESIII:2022oww, BESIII:2022vrr, BESIII:2023utd, Belle-II:2023esi, NA62:2021zjw, BESIII:2024rkp}.
However, there are fewer searches for sub-GeV invisible particles in the hadronic decays of intermediate states, where the strong interaction dominates.
The hadronic decays of intermediate states provide great potential to search for sub-GeV invisible particles, which
have no direct interactions with SM particles or with a long lifetime, such as a free gluon, ALPs coupling with gluons \cite{Ovchynnikov:2025gpx}, and other strong interaction induced particles beyond current theoretical predictions.
Searching for sub-GeV invisible particles in hadronic decays is therefore of high interest.

\begin{figure}[htb]
    \centering
    \includegraphics[width=1.0\linewidth]{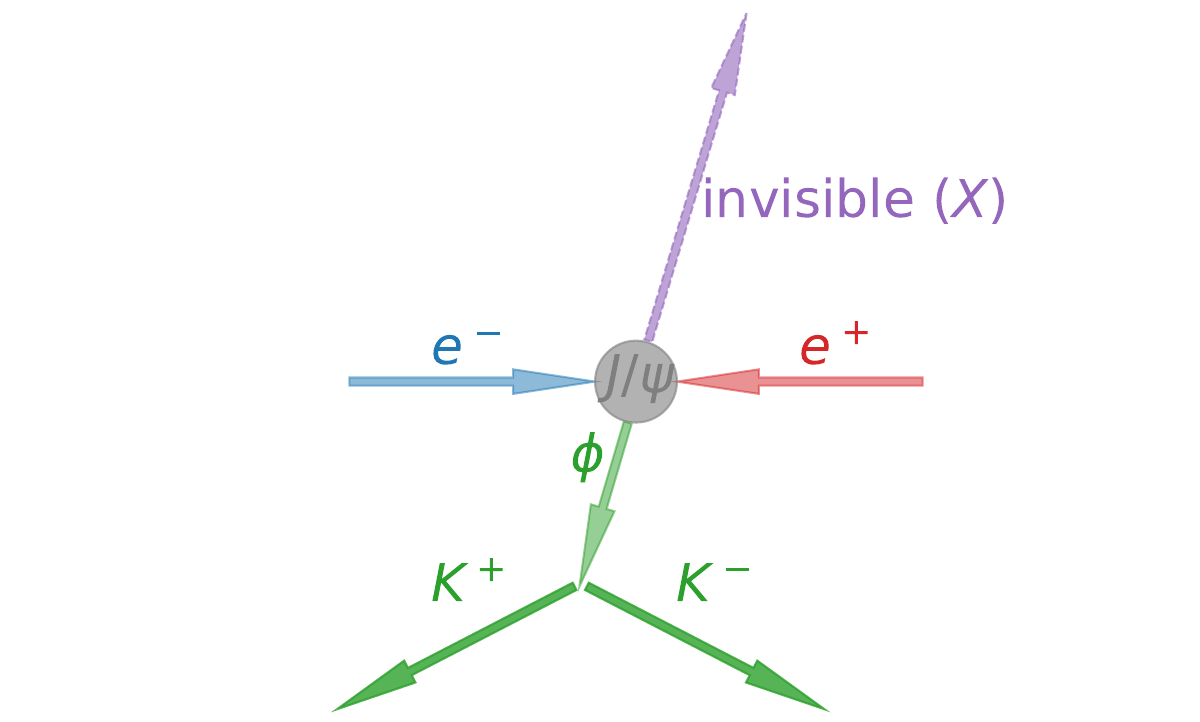}
    \caption{ The topology of $J/\psi\rightarrow\phi + X, \phi\rightarrow K^+ K^-$, where $X$ represents a sub-GeV invisible particle.}
    \label{fig:topo}
\end{figure}

In this Letter, we report a search for sub-GeV invisible particles in the process $J/\psi\rightarrow\phi + X, \phi\rightarrow K^+ K^-$ using $(8774.0\pm39.4)\times10^6$ $J/\psi$ events \cite{BESIII:2021cxx} collected by the BESIII detector from 2017 to 2019 \cite{BESIII:2009fln},
where $X$ represents a sub-GeV DM particle. 
About 10\% of the full data sample is first used to validate the analysis procedure, 
and the final result is obtained from the full data sample with the validated analysis strategy.
To veto the dominant background from the process $J/\psi\rightarrow\phi K_L^{0}K_L^0$,
the recoil mass against $\phi$ meson is required to be less than 2$m_{K^0_L}$,
where $m_{K^0_L}$ is the $K^0_L$ mass given by the Particle Data Group (PDG) ~\cite{ParticleDataGroup:2022pth}.
As shown in FIG.~\ref{fig:topo}, the simple decay topology of $\phi\rightarrow K^+K^-$ helps to suppress possible SM background processes, and provides a clean window to search for sub-GeV invisible particles.

The BESIII detector~\cite{BESIII:2009fln} records symmetric $e^+e^-$ collisions
provided by the BEPCII storage ring~\cite{Yu:2016cof}
in the center-of-mass energy ($\sqrt{s}$) range from 1.84 to 4.95~GeV,
with a peak luminosity of $1.1 \times 10^{33}\ \text{cm}^{-2}\text{s}^{-1}$
achieved at $\sqrt{s} = 3.773\ \text{GeV}$.
The cylindrical core of the BESIII detector covers 93\% of the full solid angle and consists of a helium-based multi-layer drift chamber~(MDC), a plastic scintillator time-of-flight
system~(TOF), and a CsI(Tl) electromagnetic calorimeter~(EMC),
which are all enclosed in a superconducting solenoidal magnet
providing a 1.0~T magnetic field. The solenoid is supported by an
octagonal flux-return yoke with resistive plate counter muon
identification modules interleaved with steel.
The charged-particle momentum resolution at $1~{\rm GeV}/c$ is $0.5\%$, and the
${\rm d}E/{\rm d}x$ resolution is $6\%$ for electrons from Bhabha scattering. The EMC measures photon energies with a resolution of $2.5\%$ ($5\%$) at $1$~GeV in the barrel (end cap)
region. The time resolution in the TOF barrel region is 68~ps, while
that in the end cap region is 60~ps using an upgraded multigap resistive plate chamber
technology~\cite{li2017study, guo2017study, cao2020design}.

Simulated data samples produced with a {\sc
geant4}-based~\cite{GEANT4:2002zbu} Monte Carlo (MC) package, which
includes the geometric description~\cite{Huang:2022wuo} of the BESIII detector and the
detector response, are used to determine detection efficiencies
and to estimate backgrounds. The simulation models the beam
energy spread and initial state radiation in $e^+e^-$
annihilations with the generator {\sc kkmc}~\cite{Jadach:2000ir, Jadach:1999vf}.
The inclusive MC sample simulates both the production of the $J/\psi$ resonance and the continuum processes incorporated with {\sc kkmc}~\cite{Jadach:2000ir, Jadach:1999vf}.
All particle decays are modelled with {\sc evtgen}~\cite{Lange:2001uf, Ping:2008zz} using branch fractions(BFs)
either taken from the PDG~\cite{ParticleDataGroup:2022pth}, when available,
or estimated using {\sc lundcharm}~\cite{Chen:2000tv, Yang:2014vra} if not available.
Final state radiation from charged-final-state particles is incorporated using the {\sc photos} package~\cite{Richter-Was:1992hxq}.
The signal MC samples of $J/\psi\rightarrow\phi + X, \phi\rightarrow K^+ K^-$ are generated uniformly in phase space (PHSP), where $m_{X}$ varies from 0.0 to 0.96 $\textrm{GeV}/\textit{c}^{2}$ by a step of 0.02 $\textrm{GeV}/\textit{c}^{2}$.

Charged tracks detected in the MDC are required to be within a polar angle ($\theta$) range of $|\rm{cos\theta}|<0.93$, where $\theta$ is defined with respect to the $z$-axis, the symmetry axis of the MDC.
The distance of closest approach to the interaction point must be less than 10\,cm along the $z$-axis, and less than 1\,cm in the transverse plane.
Particle identification~(PID) for charged tracks combines measurements of the energy deposited in the MDC~(d$E$/d$x$), the flight time in the TOF, and the measurements of EMC to form likelihoods $\mathcal{L}(h)~(h=p,K,\pi)$ for each hadron $h$ hypothesis.
Each charged track is assigned to the particle type that corresponds to the hypothesis with the highest confidence level (C.L.).
Photon candidates are identified using showers in the EMC.
The deposited energy of each shower must be more than 25 MeV in the barrel region ($|\rm{cos}\theta|<0.8$) and more than 50 MeV in the end cap region ($0.86<|\rm{cos}\theta|<0.92$).

\begin{figure}[htb]
    \centering
    \includegraphics[width=1\linewidth]{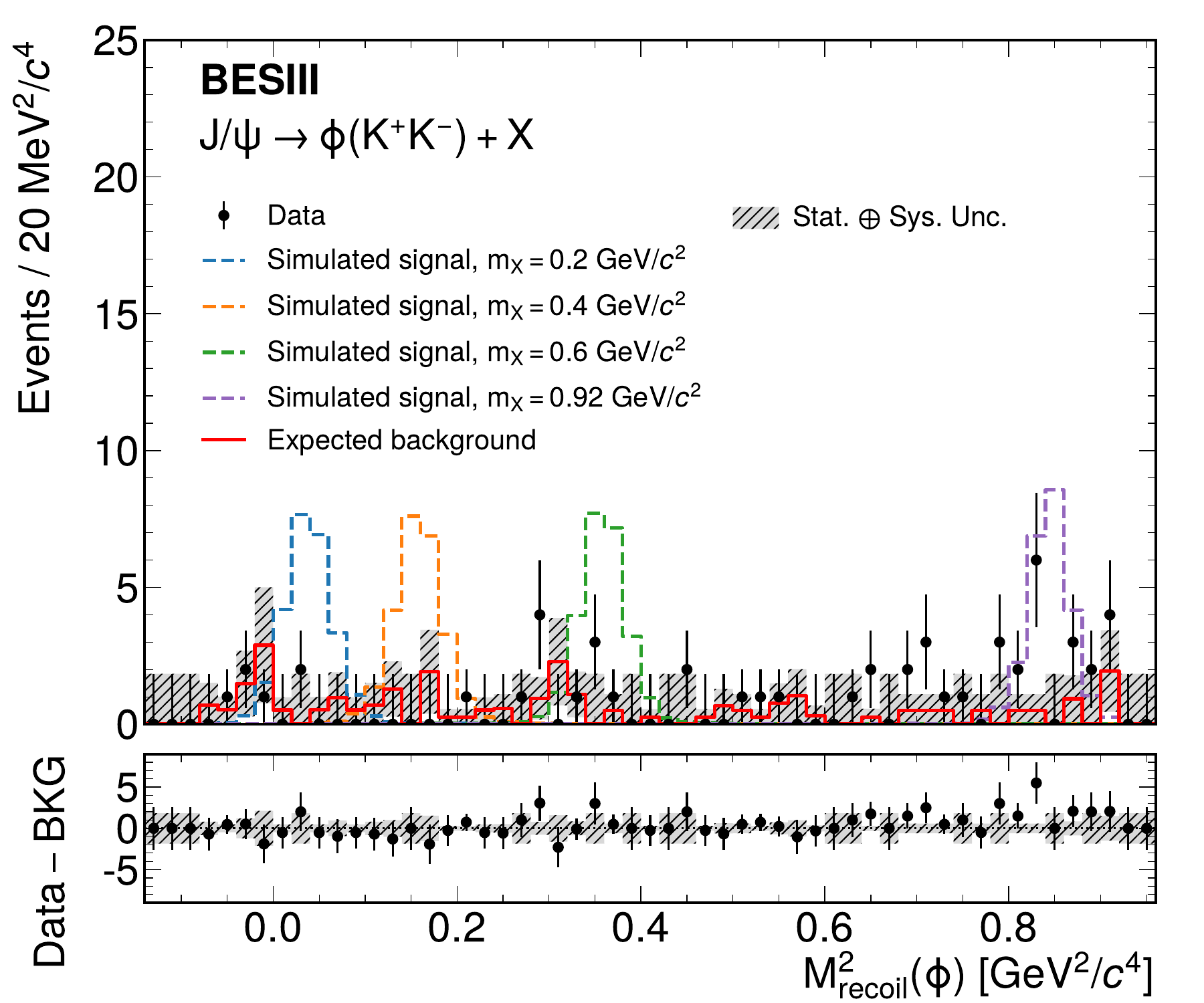}
    \caption{The distributions of $M^{2}_{\rm{recoil}}(\phi)$ for data and MC samples. The dots with error bars are the data. The blue, orange, green, and purple dashed lines are the signal MC samples with $m_{X} = $ 0.2, 0.4, 0.6, and 0.92 $\textrm{GeV}/\textit{c}^{2}$, assuming a BF of $2 \times 10^{-8}$. The red solid line is the expected background estimated by the MC study and the $e^+e^-$ collision data samples at $\sqrt{s}=$ 3.080, 3.650, and 3.682 $\textrm{GeV}$. The grey box shows the total uncertainty of the expected background, including statistical and systematic uncertainties. }
    \label{fig:recoil}
\end{figure}

Exactly 2 good charged tracks consistent with a Kaon PID hypothesis with opposite charges are required to reconstruct the $\phi$ candidate.
The invariant mass of the $K^+K^-$ candidate, $M_{K^+K^-}$, is required to satisfy $|M_{K^+K^-} - m_{\phi}| < 10\ \textrm{MeV}/\textit{c}^{2}$,  where $m_{\phi}$ is the mass of $\phi$ meson\cite{ParticleDataGroup:2022pth}.
Events with additional charged tracks are excluded from further analysis.
The recoil mass against the $\phi$ candidate, $M_{\rm{recoil}}(\phi) = \sqrt{(p_{e^+e^-} - p_{K^+} - p_{K^-})^{2}}$, is used to suppress the background from $J/\psi\rightarrow\phi K_L^0 K_L^0$ with the requirement $M_{\rm{recoil}}(\phi) < 0.96\ \textrm{GeV}/\textit{c}^{2}$,
where $p_{e^+e^-}$ \cite{BESIII:2021cxx}, $p_{K^+}$, and $p_{K^-}$ are the four-momenta of the initial $e^+e^-$ system, the $K^+$ meson, and the $K^-$ meson, respectively.
To reduce the backgrounds caused by $J/\psi\rightarrow\phi\eta$ and $J/\psi\rightarrow\phi\pi^{0}$, the number of good photons, $N_\gamma$, is required to be 0 in the candidate event~\cite{Li:2024pox}.
To further veto the backgrounds from $J/\psi\rightarrow\phi\eta$ and $J/\psi\rightarrow\gamma\eta_{c}$, the polar angle of the recoil momentum against $\phi$, $\theta_{\rm{recoil}}(\phi)$, is required to point within the central barrel EMC region, $|\rm{cos}\theta_{\rm{recoil}}(\phi)| < 0.7$.
To suppress $J/\psi\rightarrow\pi^{\pm}K^{\mp}K_L^0$ background originating from $\pi$-$K$ misidentification, events with $M_{\textrm{recoil}}(\pi^{\pm}K^{\mp}) < 0.56\ \textrm{GeV}/\textit{c}^{2}$ are excluded.
This is achieved by recalculating $M_{\textrm{recoil}}(\pi^{\pm}K^{\mp})$ through the switching of one charged track's mass hypothesis from Kaon to Pion.
The thresholds for the relevant event selection criteria are determined by maximizing  $n_s/\sqrt{n_s+n_b}$, where the signal yield $n_s$ is extracted from signal MC samples, the background yield $n_b$ is determined from the data, and the BF of signal MC is assumed to be $\sim10^{-6}$.
As a check, the Punzi figure of merit~\cite{Punzi:2003bu} with 3$\sigma$ target significance yields similar results on these event selection thresholds.
After the above selection criteria, 52 events remain in the distribution of $M^{2}_{\rm{recoil}}(\phi)$ as shown in FIG.~\ref{fig:recoil}, where the Kaon momentum resolution and the beam energy measurement resolution can result in negative values of $M_{\textrm{recoil}}^2$. 
Additionally, no significant peak is seen in the searched mass region.

Background events are separated into two categories: the $J/\psi$ decay processes and the $e^+e^-$ continuum processes.
The background contributions from the $J/\psi$ decay processes are studied using the inclusive MC sample, the exclusive MC samples generated according to the survived processes of the inclusive MC sample.
The sum of the normalized exclusive MC samples is treated as the expected background from the $J/\psi$ decay,
where the exclusive MC samples are normalized to the luminosity of the data \cite{BESIII:2021cxx} using associated BFs reported by the PDG \cite{ParticleDataGroup:2022pth}.
Meanwhile, the background contributions from the $e^+e^-$ continuum processes are investigated with the $e^+e^-$ collision data samples at $\sqrt{s}=$ 3.080, 3.650, and 3.682 $\textrm{GeV}$ corresponding to integrated luminosities of 136, 401, 395 $\rm{pb}^{-1}$.
The expected background from the $e^+e^-$ continuum processes is estimated by a data-driven method, where the $e^+e^-$ collision data samples are employed to determine the possible backgrounds by \cite{BESIII:2021cxx}
\begin{equation}
    \begin{aligned}
    N_{\rm{bkg}} = N_{i} \times \frac{\mathcal{L}_{J/\psi}}{\mathcal{L}_{i}} \times \frac{s_i}{s_{J/\psi}}.
    \end{aligned}
    \label{eq:bkg}
\end{equation}
The $i$ indicates the $e^+e^-$ collision data at $\sqrt{s}=$ 3.080, 3.650, and 3.682 $\textrm{GeV}$; $N_{\rm{bkg}}$ is the number of expected background events from the $e^+e^-$ continuum processes;
$N_i$ is the number of survived events for the $i$ data;
$\mathcal{L}_{J/\psi}$ and $\mathcal{L}_{i}$ are the integrated luminosities for the $J/\psi$ and data $i$;
$s_{J/\psi}$ and $s_i$ are the corresponding squares of the center-of-mass energies.
The $N_{\rm{bkg}}$ is calculated to be ($0.0\pm1.8$) using the $e^+e^-$ collision data collected at $\sqrt{s}=$ 3.080, 3.650, and 3.682 $\textrm{GeV}$.
The expected yields of different background sources are summarized in TABLE~\ref{tab:bkg}.
\begin{table}[!htbp]
    \centering
    \caption{The expected yields of different background sources, where only the statistical uncertainties from the size of MC samples are included.}
    \begin{tabular}{cc}
        \hline
        \hline
        Source & Yield \\
        \hline
         $J/\psi\rightarrow K^{*0}\overline{K}^{0},\ K^{*0}\rightarrow\pi^- K^+,\ \overline{K}^0\rightarrow K_L^0$ & $6.3\pm2.2$ \\
         $J/\psi\rightarrow\overline{K}^{*0}{K}^{0},\ \overline{K}^{*0}\rightarrow\pi^+ K^-,\ {K}^0\rightarrow K_L^0$ & $0.7\pm0.7$ \\
         $J/\psi\rightarrow\pi^{0}\overline{K^{0}}{K}^{*0},\overline{K^0}\rightarrow K_L^0, {K}^{*0}\rightarrow\pi^-K^+$ & $0.3\pm0.4$ \\
         $J/\psi\rightarrow\pi^{+}K^{-}{K}^{*,0},\ {K}^{*,0}\rightarrow K^0\pi^0, K^0\rightarrow K_L^0$ & $0.4\pm0.4$ \\
         $J/\psi\rightarrow\pi^{+}K^{-}\bar{K}^0(\textrm{PHSP}),\ \bar{K}^0\rightarrow K_L^0$ & $1.3\pm1.9$ \\
         $J/\psi\rightarrow\pi^{-}K^{+}{K}^0(\textrm{PHSP}),\ {K}^0\rightarrow K_L^0$ & $1.1\pm1.5$ \\
         $J/\psi\rightarrow\phi K^0\overline{K^0}, K^0\rightarrow K_L^0, \overline{K^0}\rightarrow K^0_L$ & $7.6\pm2.7$ \\
         $J/\psi\rightarrow\phi\eta^{\prime},\eta^{\prime}\rightarrow\ \rm{anything}$ & $0.9\pm0.6$ \\
         $J/\psi\rightarrow\phi\eta,\eta\rightarrow\ \rm{anything}$ & $3.2\pm1.0$ \\
         $J/\psi\rightarrow\gamma\eta_c,\ \eta_c\rightarrow\pi^+ K^0 K^-,\ K^0\rightarrow K_L^0$ & $2.8\pm0.9$\\
         $J/\psi\rightarrow\gamma\eta_c,\ \eta_c\rightarrow\pi^- \overline{K^0} K^+,\ \overline{K^0}\rightarrow K_L^0$ & $4.1\pm1.0$\\
         $e^+e^-$ continuum data & $0.0\pm1.8$\\
         \hline
         Total & $28.7\pm 5.0$ \\
         \hline
         \hline
    \end{tabular}
    \label{tab:bkg}
  \end{table}

The systematic uncertainty in the expected background yield, originated from the background estimation method, is derived from the difference between the expected background and data in the control region $1.04<M_{K^+K^-}<1.07\ \textrm{GeV}/c^2$, quantified as $57.0\%$.
In addition, the systematic uncertainties of the expected background yield are from charged track detection and PID, $|M_{K^+K^-}-m_{\phi}| < 10\ \textrm{MeV}/\textit{c}^{2}$, $N_\gamma = 0$, $|\cos\theta_{\rm{recoil}}(\phi)| < 0.7$, and $M_{\rm{recoil}}(\pi^{\pm}K^{\mp})<0.56\ \textrm{GeV}/\textit{c}^{2}$.
The systematic uncertainties due to charged track detection and PID in the expected background yield are assigned as $2\%$ and $2\%$ \cite{BESIII:2016tbd}, respectively.
The systematic uncertainty in the expected background yield caused by the $\phi$ mass window is determined to be $2.9\%$ by varying it with $\pm 1\sigma$.
The systematic uncertainty relevant to $N_\gamma = 0$ is estimated to be $9.8\%$ in the expected background yield, as determined by the discrepancy between data and MC using the control sample of $J/\psi\rightarrow \mu^+\mu^-$.
The systematic uncertainty associated with $|\cos\theta_{\rm{recoil}}(\phi)| < 0.7$ is estimated to be 0.2\% using the control sample of $J/\psi\rightarrow\phi\eta,\eta\rightarrow\gamma\gamma$.
The systematic uncertainty of $M_{\rm{recoil}}(\pi^{\pm}K^{\mp})<0.56\ \textrm{GeV}/\textit{c}^{2}$ is estimated to be $2.3\%$ with the control sample of $J/\psi\rightarrow \pi^{\pm}K^{\mp}K_S^0, K^0_S\rightarrow\pi^+\pi^-$.
Assuming all these sources are independent, the total systematic uncertainty of the expected background yield is retrieved as the quadrature sum of the individual contributions. The expected background yield is then determined to ($28.7\pm5.0_{\rm{stat.}}\pm16.6_{\rm{sys.}}$), which is consistent with the data yield within 1.5$\sigma$, as shown in FIG.~\ref{fig:recoil}.

A profile likelihood method \cite{Cowan:2010js} is used to extract the upper limit on the number of $J/\psi\rightarrow\phi + X$ events at 90\% C.L..
In the procedure, the data is assumed to obey a Poisson distribution, while the expected background and the signal detection efficiency are treated as Gaussian distributions.
The likelihood function is calculated as
\begin{equation}
    \begin{aligned}
    \mathcal{L} = & \prod \mathcal{F}(x,\ y,\ z | \ \mu,\ b,\ \epsilon) \\
    = & \prod \frac{(\epsilon\mu+b)^{x}}{x!}e^{-(\epsilon\mu+b)} \cdot \frac{1}{\sqrt{2\pi}\sigma_b} e^{-\frac{(y-b)^2}{2\sigma_b^2}} \cdot \frac{1}{\sqrt{2\pi}\sigma_\epsilon} e^{-\frac{(z-\epsilon)^2}{2\sigma_\epsilon^2}},
    \end{aligned}
    \label{eq:likeli}
\end{equation}
where $x$ is the number of observed data events, $y$ is the background yield,
$z$ is the signal detection efficiency,
$\mu$ is the expected signal yield,
$b$ is the expected background yield,
$\sigma_{b}$ is the uncertainty of $b$,
$\epsilon$ is the expected signal detection efficiency, and $\sigma_{\epsilon}$ is the uncertainty of $\epsilon$.

The signal detection efficiency is determined using the signal MC samples of $J/\psi\rightarrow\phi + X, \phi\rightarrow K^+ K^-$,
with $m_{X}$ varies from 0.0 to 0.96 $\textrm{GeV}/\textit{c}^{2}$ by a step of 0.02 $\textrm{GeV}/\textit{c}^{2}$.
The detection efficiencies vary from 18.0\% to 19.8\% for the different signal MC samples, and the average value of the detection efficiencies estimated using the different signal MC samples, $(19.27\pm0.03_{\rm{stat.}})\%$, is taken as the signal detection efficiency.

The sources of systematic uncertainties of the signal detection efficiency come mainly from the same sources as the expected background yield except the background estimation method.
The systematic uncertainties in the signal detection efficiency are determined to be $0.4\%$, $0.4\%$, $2.9\%$, $1.7\%$, $0.04\%$, and $0.4\%$ related to the same sources mentioned in the systematic uncertainty of expected background yield, respectively.
Moreover, the difference between the alternative and nominal detection efficiencies, 2.1\%, is taken as the systematic uncertainty caused by the signal model.
Alternative detection efficiencies are obtained from signal MC samples that were generated using a different angular distribution.
Assuming that all these sources are independent, the total systematic uncertainty of the signal detection efficiency is taken as the quadratic sum of the individual contributions,
resulting in 4.0\%.
The final signal detection efficiency is determined to be $(19.27\pm0.03_{\rm{stat.}}\pm4.02_{\rm{sys.}})\%$, which enters the upper limit calculation.

\begin{figure}[htb]
    \centering
    \includegraphics[width=0.9\linewidth]{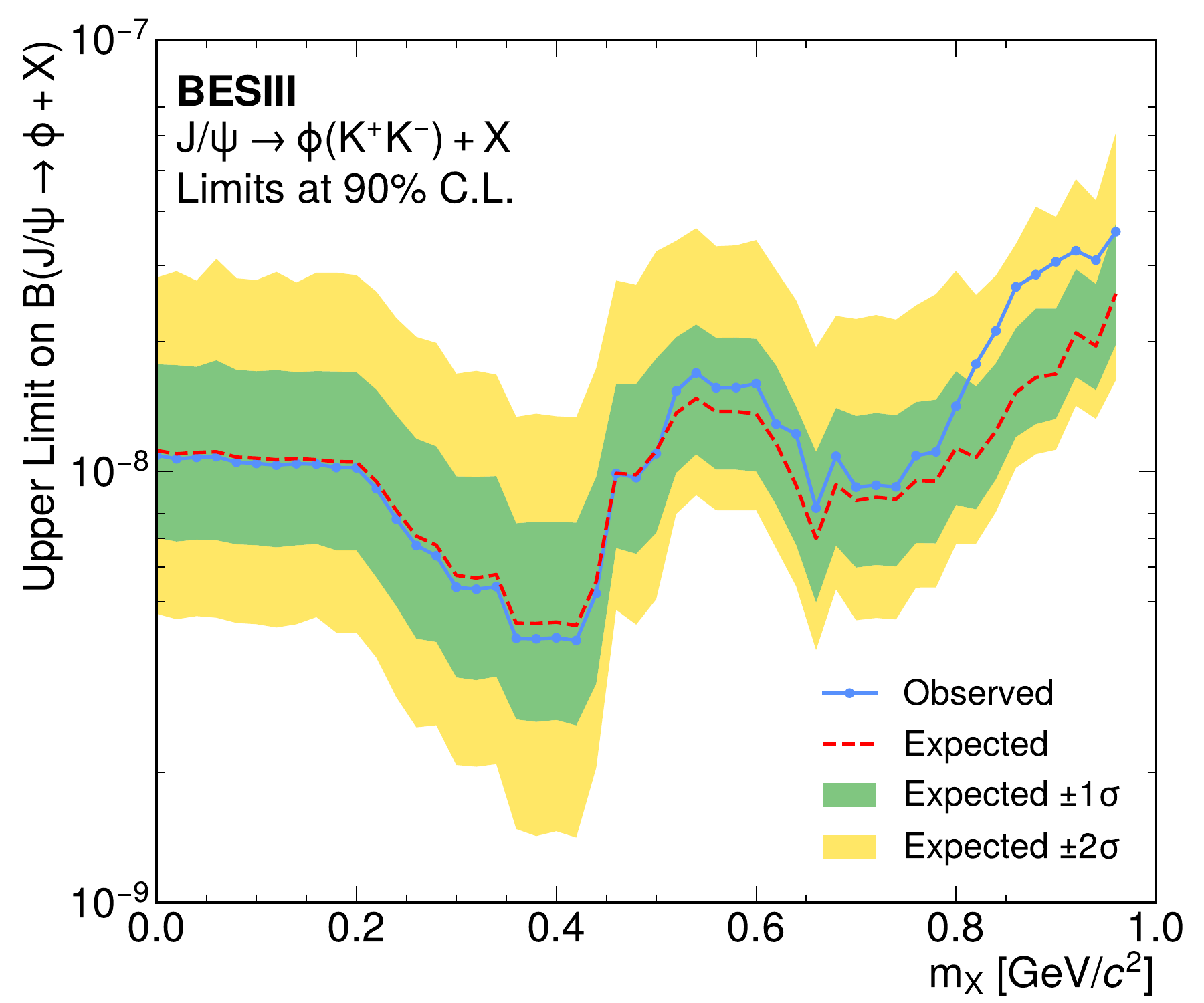}
    \caption{The observed (blue points) and expected (red dashed line) upper limits on $\mathcal{B}(J/\psi\rightarrow\phi + X)$ as a function of $m_{X}$ in the accessible $m_{X}$ range at 90\% C.L.. The expected upper limits and their uncertainties are obtained using the background model with no signal. The $\pm1\sigma$ and $\pm2\sigma$ expected upper limit bands are shown as the green and yellow bands, respectively. }
    \label{fig:up}
\end{figure}

The upper limit on the BF of $J/\psi\rightarrow\phi + X$ at 90\% C.L. is calculated by
\begin{equation}
    \begin{aligned}
    \mathcal{B}(J/\psi\rightarrow\phi + X) = \frac{N_{X}}{N_{J/\psi} \cdot \mathcal{B}(\phi\rightarrow K^+K^-)},
    \end{aligned}
    \label{eq:up}
\end{equation}
where $N_{X}$ is the 90\% C.L. upper limit on the signal yield after efficiency correction,
$N_{J/\psi} = (8774.0\pm39.4)\times10^6$ is the total number of $J/\psi$ events in data,
and $\mathcal{B}(\phi\rightarrow K^+K^-)$ is the BF of $\phi\rightarrow K^+K^-$ \cite{ParticleDataGroup:2022pth}.
The upper limit on the inclusive BF of $J/\psi\rightarrow\phi + X$ is determined
to be $7.0\times 10^{-8}$ at 90\% C.L. in the region of $0.0 < m_{X} < 0.96\ \textrm{GeV}/\textit{c}^{2}$, where the related numbers are summarized in TABLE~\ref{tab:result}.
\begin{table}[!htbp]
    \centering
    \caption{Summary of the number of events observed in data $N_{\rm{observed}}$, the expected number of background events $N_{\rm{background}}$, the detection efficiency $\epsilon$, and the upper limit on signal yield at 90\% C.L., $N_{X}$. }
    \begin{tabular}{ccccc}
        \hline
        \hline
        Process & $N_{\rm{observed}}$ & $N_{\rm{background}}$ & $\epsilon$ [\%] & $N_{X}$ \\
        \hline
         $J/\psi\rightarrow\phi + \rm{invisible}$ & $52$ & $28.7\pm17.0$ & $19.3\pm4.0$ & $303.3$ \\
         $\eta\rightarrow\rm{invisible}$ & $11$ & $5.9\pm3.3$ & $17.5\pm3.4$ & $76.9$ \\
         \hline
         \hline
    \end{tabular}
    \label{tab:result}
\end{table}
The 90\% C.L. upper limits on $\mathcal{B}(J/\psi\rightarrow\phi + X)$ are extracted as a function of $m_{X}$,
where the signal detection efficiencies are calculated from a set of signal MC samples with different $m_{X}$.
In addition to the above selection criteria, a signal window of $|M_{\rm{recoil}}^2(\phi) - m_{X}^2| < 90\ \rm{MeV}^2/\textit{c}^{4}$ is applied, corresponding to approximately 5 times the resolution of $M_{\rm{recoil}}^2(\phi)$.
The 90\% C.L. upper limits on $\mathcal{B}(J/\psi\rightarrow\phi + X)$ as a function of $m_{X}$ vary from  $4\times10^{-9}$ to $4\times10^{-8}$ in the accessible $m_{X}$ range, as shown in FIG.~\ref{fig:up}. This is the first determination of the upper limits on $\mathcal{B}(J/\psi\rightarrow\phi + X)$ in the sub-GeV region.
Following the model-dependent theoretical framework of a scalar mediator portal \cite{Boveia:2016mrp}, this work has comparable sensitivity to the recent sub-GeV DM studies \cite{SENSEI:2023zdf} in the sub-GeV DM mass range from 0.5 to 0.9 $\textrm{GeV}/\textit{c}^{2}$.
Notably, the interpretation approach is developed for LHC experiments and not totally suitable in this work.
A robust and systematic approach is essential to effectively compare these findings with other sub-GeV DM detection results \cite{SENSEI:2023zdf, XENON:2019zpr, Essig:2019xkx, DarkSide:2022knj}.

In addition,
the BF of $\eta\rightarrow\rm{invisible}$ is also determined as
\begin{equation}
    \begin{aligned}
    \mathcal{B}(\eta\rightarrow{\rm{invisible}}) = \frac{N_{\eta\rightarrow{\rm{invisible}}}}{N_{J/\psi}\cdot \mathcal{B}(J/\psi\rightarrow\phi\eta) \cdot \mathcal{B}(\phi\rightarrow K^+K^-)},
    \end{aligned}
    \label{eq:up_eta}
\end{equation}
where $N_{\eta\rightarrow{\rm{invisible}}}$ is the 90\% C.L. upper limit on the signal yield of $\eta\rightarrow{\rm{invisible}}$ after efficiency correction, $\mathcal{B}(J/\psi\rightarrow\phi\eta)$ \cite{ParticleDataGroup:2022pth} is the corresponding BF of $J/\psi\rightarrow\phi\eta$.
A signal window of  $|M^{2}_{\rm{recoil}}(\phi) - m_{\eta}^2| < 90\ \rm{MeV}^2/\textit{c}^{4}$, is employed to calculate the upper limit on the signal yield of $J/\psi\rightarrow\phi\eta, \eta\rightarrow\rm{invisible}$.
Using the signal MC sample of $J/\psi\rightarrow\phi\eta, \eta\rightarrow\rm{invisible}$, the selection efficiency is determined as $(17.5\pm0.2_{\textrm{stat.}}\pm3.4_{\textrm{sys.}})\%$, as shown in TABLE~\ref{tab:result}.
The systematic uncertainties are estimated by the same ways in the process $J/\psi\rightarrow\phi + X, \phi\rightarrow K^+K^-$.
Using Eq.~\ref{eq:up_eta}, the upper limit on $\mathcal{B}(\eta\rightarrow{\rm{invisible}})$ at 90\% C.L. is determined as $2.4\times10^{-5}$, which improves the previous
best results \cite{BESIII:2012nen,NA64:2024azv} by more than four times.

In summary, a search for sub-GeV invisible particles in the process $J/\psi\rightarrow\phi + X, \phi\rightarrow K^+ K^-$ is performed for the first time using $(8774.0\pm39.4)\times 10^6$ $J/\psi$ events collected by the BESIII detector.
No significant signal over the expected background is observed in the accessible invisible particle mass region $m_{X} < 0.96\ \textrm{GeV}/\textit{c}^{2}$,
and the upper limit on the inclusive BF of $J/\psi\rightarrow\phi + X$ is determined to be $7.0\times10^{-8}$ at 90\% C.L.
The upper limits on this BF are also calculated as a function of $m_{X}$, which vary from $4\times10^{-9}$ to $4\times10^{-8}$ in the same search region.
Moreover, the 90\% C.L. upper limit on the BF of $\eta\rightarrow \rm{invisible}$ at is set as $2.4\times10^{-5}$, which is more stringent than the previous
best results \cite{BESIII:2012nen,NA64:2024azv}.
This study shows the great potential for searching sub-GeV invisible particles in the hadronic decays of $J/\psi$, which could significantly improve the search for sub-GeV invisible particles involving the strong interaction, such as a free gluon, ALPs coupling with gluons, and possible strong interaction induced invisible particles.

\acknowledgments

The BESIII Collaboration thanks the staff of BEPCII (https://cstr.cn/31109.02.BEPC) and the IHEP computing center for their strong support. This work is supported in part by National Key R\&D Program of China under Contracts Nos. 2023YFA1606000, 2023YFA1606704, 2020YFA0406400, 2020YFA0406300; National Natural Science Foundation of China (NSFC) under Contracts Nos. 11635010, 11735014, 11935015, 11935016, 11935018, 12025502, 12035009, 12035013, 12061131003, 12192260, 12192261, 12192262, 12192263, 12192264, 12192265, 12221005, 12225509, 12235017, 12361141819; the Chinese Academy of Sciences (CAS) Large-Scale Scientific Facility Program; the CAS Center for Excellence in Particle Physics (CCEPP); Joint Large-Scale Scientific Facility Funds of the NSFC and CAS under Contract No. U1832207; CAS under Contract No. YSBR-101; 100 Talents Program of CAS; The Institute of Nuclear and Particle Physics (INPAC) and Shanghai Key Laboratory for Particle Physics and Cosmology; Agencia Nacional de Investigación y Desarrollo de Chile (ANID), Chile under Contract No. ANID PIA/APOYO AFB230003; German Research Foundation DFG under Contract No. FOR5327; Istituto Nazionale di Fisica Nucleare, Italy; Knut and Alice Wallenberg Foundation under Contracts Nos. 2021.0174, 2021.0299; Ministry of Development of Turkey under Contract No. DPT2006K-120470; National Research Foundation of Korea under Contract No. NRF-2022R1A2C1092335; National Science and Technology fund of Mongolia; National Science Research and Innovation Fund (NSRF) via the Program Management Unit for Human Resources \& Institutional Development, Research and Innovation of Thailand under Contract No. B50G670107; Polish National Science Centre under Contract No. 2019/35/O/ST2/02907; Swedish Research Council under Contract No. 2019.04595; The Swedish Foundation for International Cooperation in Research and Higher Education under Contract No. CH2018-7756; U. S. Department of Energy under Contract No. DE-FG02-05ER41374.
\newline

%\nocite{*}

\bibliography{output}% Produces the bibliography via BibTeX.

%
% ****** End of file apssamp.tex ******

\end{document}